\documentclass[12pt]{iopart}
\usepackage{iopams}
\usepackage{xspace}
\usepackage{graphicx}
\usepackage{epsfig}
\usepackage{lineno}

\begin{document}

\title[CAST-TPC background studies and shielding effects]{Background studies and shielding effects for the TPC detector of the
CAST experiment}

\author{G Luz\'{o}n, B Beltr\'{a}n\footnote{Present address: Department of Physics, Queens University, Kingston, Ontario, Canada}, J M Carmona, S Cebri\'{a}n, H G\'{o}mez, I~G~Irastorza, J
Morales, A Ort\'{\i}z, A Rodr\'{\i}guez, J Ruz, J~A~Villar}
\address{Instituto de F\'{\i}sica Nuclear y Altas Energ\'{\i}as, Universidad de Zaragoza, Zaragoza,
Spain} \ead{luzon@unizar.es}

\begin{abstract}
Sunset solar axions traversing the intense magnetic field of the
CERN Axion Solar Telescope (CAST) experiment may be detected in a
Time Projection Chamber (TPC) detector, as point-like X-rays
signals. These signals could be masked, however, by the
inhomogeneous  background of materials in the experimental site. A
detailed analysis, based on the detector
characteristics, the background radiation at the CAST site, simulations
and experimental results, has allowed us to design a shielding which
reduces  the background level by a factor of $\sim$4 compared to the
detector without shielding, depending on its position, in the energy range between 1 and
$10\,\rm{keV}$. Moreover, this shielding has improved the homogeneity
of background measured by the TPC.

\end{abstract}

\pacs{29.40.Cs, 95.35.+d, 07.85.Nc, 07.05.Fb, 07.05.Kf}

 \maketitle
 \section{Introduction \label{sec:intro}}

The CAST experiment \cite{CAST1} is placed at CERN and makes use of a
decommissioned LHC test magnet to look for solar axions through its
Primakoff conversion into photons inside the magnetic field. The 10\,m-length
magnet is installed
on a platform which allows it to move $\pm 8^\circ$ vertically and
$\pm 100^\circ$ horizontally to track the sun during the sunset and
the sunrise. The 9 Tesla magnetic field is confined in two parallel pipes of 4.2
 cm of diameter. At the end of the pipes, three detectors  look for the X-rays
originated by the conversion of the axions inside the magnet when it
points to the Sun. The two apertures of one of the ends of the
magnet are covered by a conventional Time Projection Chamber (TPC)
\cite{CASTTPC} facing ``sunset" axions while in the opposite end,
 a Charge Coupled Device (CCD)
\cite{CCDT} coupled to an X-ray focusing telescope \cite{teles}, and
a Micromegas detector \cite{MMT} search for ``sunrise" axions. The
first results from the 2003 data analysis implied an upper limit to
the axion-photon coupling $g_{a\gamma}<1.16\times 10^{-10}$
GeV$^{-1}$ \cite{Castpaper}. A second set of measurements
corresponding to 2004 with improved detectors and longer exposure
set an upper limit on the axion-photon coupling of
$g_{a\gamma}<8.8\times 10^{-11}\,\rm{GeV}^{-1}$
 at 95$\%$ CL for axion masses $\le 0.02\,\rm{eV}$ \cite{CAST2}.

All  three detectors use discrimination techniques to reduce
background contamination of the expected signal. The X-ray signal
produced by the axions inside the magnet has a maximum at 4\,keV and vanishes at around
10\,keV \cite{CAST2}. More energetic deposits of energy, or
signals outside the detector volume facing the apertures of the magnet,
should be rejected. Simple requirements, depending on each detector,
eliminate most of the background due to charged particles like
cosmic rays, alpha and beta radiation \cite{CASTTPC,CCDT,teles,MMT}.

After the use of software cuts, the main source of background is
expected to be gamma rays
 produced predominantly in the radioactive chains of $^{238}$U,
$^{232}$Th and in the $^{40}$K isotope decays. Their interactions
with material near the detectors can generate low energy photons via the
Compton effect and also X-rays (as those from copper identified in
experimental spectra \cite{CCDT}). Additional sources are neutrons
produced by fission and ($\alpha$,n) processes or those induced by
muons and cosmic rays. As the detector is moving tracking the sun,
the background, and its inhomogeneity, is a source of uncertainty
since the axion signal is computed as the tracking (magnet pointing
to the sun) minus the background (any other magnet orientation)
signal. Therefore, not only low
background levels but also a certain independence of the position
are desirable to reduce uncertainties. The shielding presented here has been built for the
TPC with this aim. In this article we will study the effects of the complete shielding installed in 2004, comparing its
performance to the copper box used as shielding during the 2003 data taking
period. For these studies, the energy range 3--7\,keV has been chosen as the control
region to analyse in order to be sure to eliminate threshold or
saturation effects. Preliminary results were presented in
\cite{TAUP}.

In the first Section, we will describe briefly the detector and the
shielding. Next, the experimental site and its background will be
analyzed in detail taking into account gamma and neutron sources.
Monte Carlo simulations of the different sources of background will
be discussed in Section~\ref{sec:simulations}. In
Section~\ref{sec:tpc-measurements},
 the effects of the shielding on background data will be
 studied. Finally, a few remarks summarizing the
study.

\section{The shielding of the CAST TPC detector \label{sec:tpc}}

The CAST TPC detector has a conversion volume of $10\times 15 \times
30\,\rm{cm}^3$ filled with Ar(95\%)-CH$_4$(5\%) gas at atmospheric
pressure. The 10\,cm drift direction is parallel to the magnet beam axis  and
perpendicular to the section of $15 \times 30\,\rm{cm}^2$ covering
both magnet apertures. Two $6\,\rm{cm}$ diameter windows, consisting of
very thin mylar foils (3 or $5\,\mu\rm{m}$) stretched on a metallic
strongback, allow the X-rays coming from the magnet to enter the
chamber \cite{CASTTPC}. Except for the electrodes, the screws, the Printed Circuit
Board (PCB) and  the windows, the entire chamber is made of 1.7\,cm
thick low radioactivity plexiglass.

A shielding around the TPC was designed to reduce the background in
a complementary way to the effect of the off-line software cuts
\cite{CASTTPC}. The requirements were to get a certain reduction of
the background levels coming from external sources and  a decrease
of the background spatial inhomogeneity observed in the experimental site. The final
design has been the result of a compromise among the shielding
effect and technical limitations such as weight and size restrictions imposed by the experimental moving structure.

\begin{figure}[ht]
\begin{center}

 \hspace{0 cm}
\includegraphics[width=70mm]{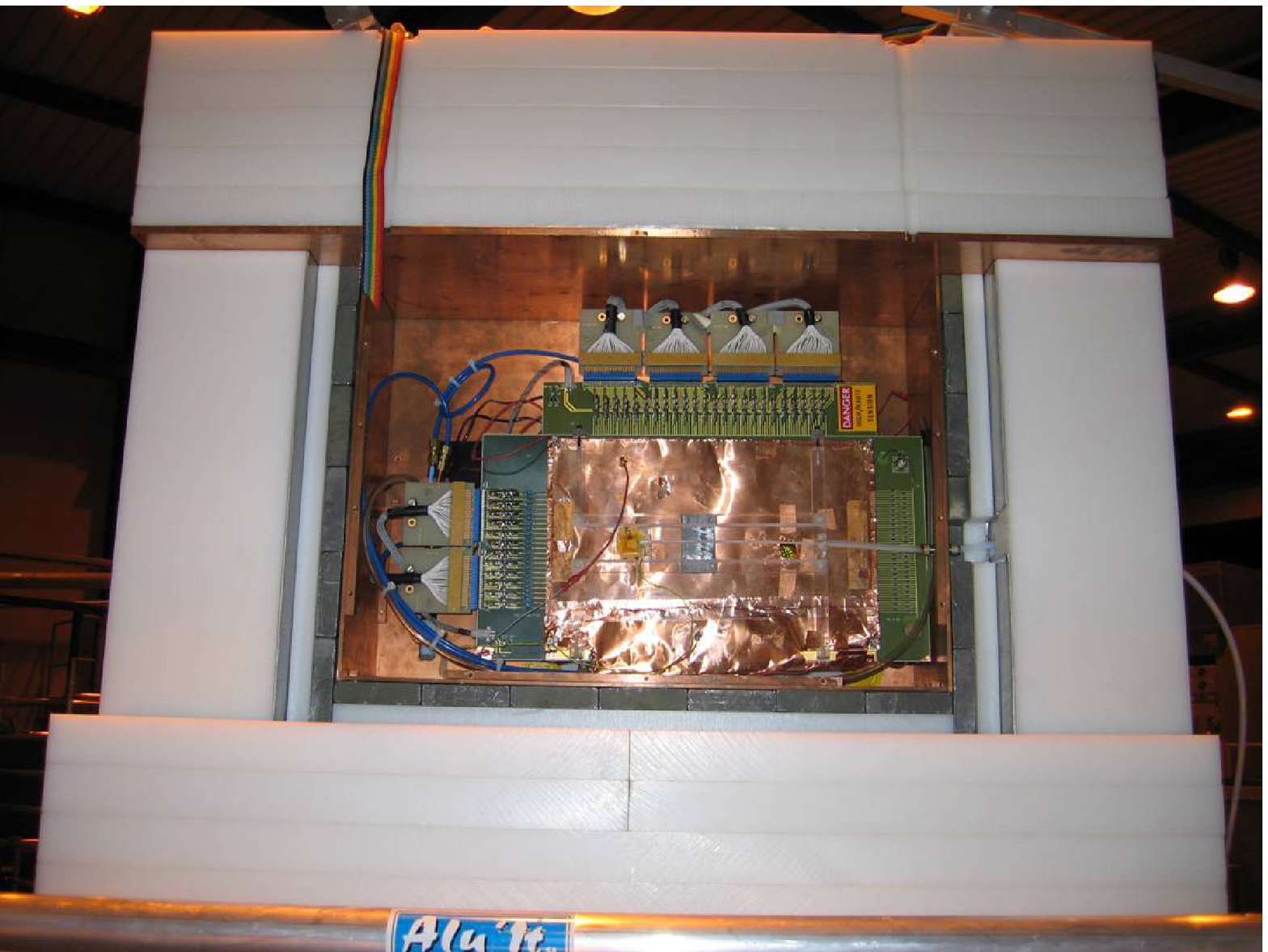}
\includegraphics[width=70mm]{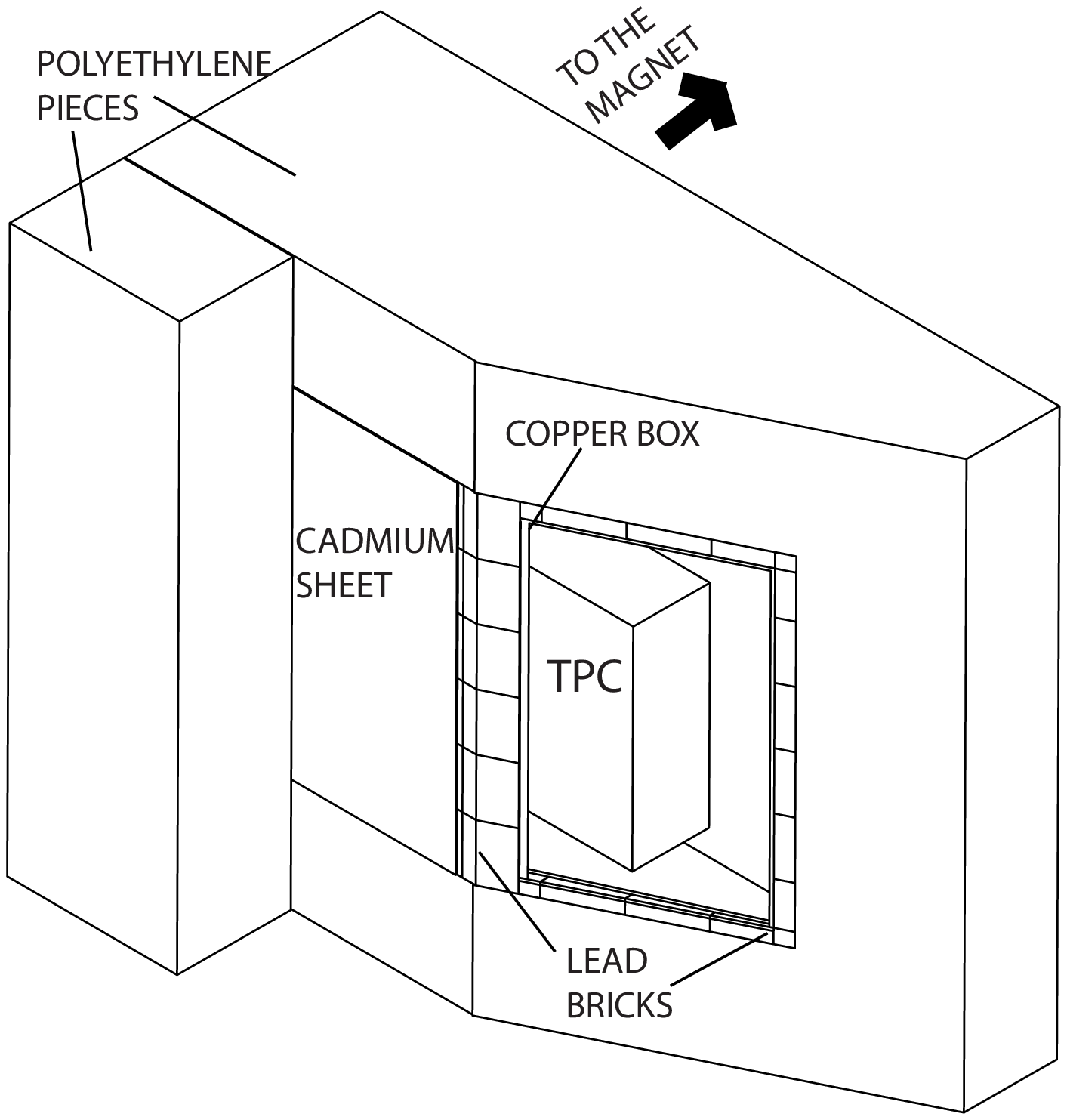}
\caption{\label{shielding}Picture on the left: TPC chamber inside
the open shielding, consisting of: a copper box and lead,
cadmium and polyethylene shields. A scheme has been drawn on the
right to show the 3D arrangement of the whole structure.}
\end{center}
\end{figure}

From inside to outside, the CAST TPC shielding (see Figure~\ref{shielding}) is composed of:
\begin{itemize}
\item[-]A copper box, $5\,\rm{mm}$ thick. This Faraday cage
reduces
  the electronic noise and stops low energy X-rays produced in the
  outer part of the shielding by environmental gamma radiation. It is also used for
mechanical support purposes.
  \item [-]Lead bricks, $2.5\,\rm{cm}$ thick, which reduces the low and medium energy environmental gamma
  radiation.
  \item [-]A cadmium layer, $1\,\rm{mm}$ thick, to absorb the thermal neutrons slowed down
  by the outer polyethylene wall.
  \item [-]Polyethylene pieces, $22.5\,\rm{cm}$ thick, used to slow the medium energy
  environmental neutrons down to thermal energies. It also reduces
  the gamma contamination and helps the mechanical stability of
  the whole structure.
  \item [-]A PVC bag which covers the whole shielding assembly. This
  tightly closes the entire set-up allowing us to flush the inner part with pure
  N$_2$ gas coming from liquid nitrogen evaporation in order to purge this space
  of radon.
  \item [-]A scintillating veto, $80\times40\times 5\,\rm{cm}^3$, placed at the top of the shielding to reject
muon-induced events working in anti-coincidence with the detector.
\end{itemize}

 The described scheme is the
outcome of several simulations  and experimental tests. During the
2003 data taking period, the detector set-up consisted just of a
copper box with  N$_2$ gas flush. The full shielding was installed
in 2004, after a test carried out in the real experimental
conditions.

\section{Experimental site and background\label{sec:background}}

\begin{figure}[ht]
\begin{center}

 \hspace{0 cm}
\includegraphics[width=120mm]{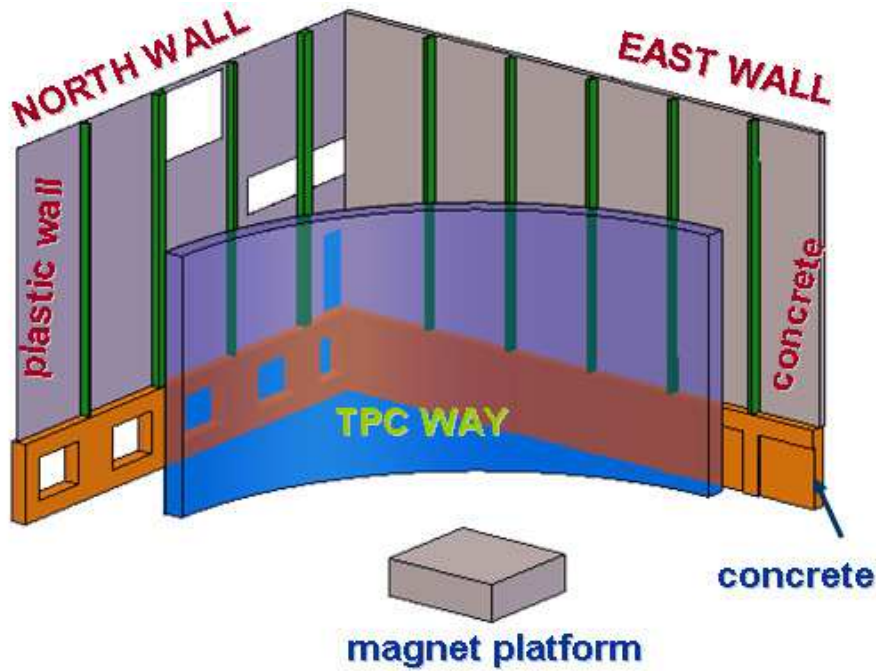}
\caption{\label{Site} Graph shows the TPC detector path facing the North and East walls.}
\end{center}
\end{figure}

The CAST experiment is located at one of the buildings of the SR8
experimental area at CERN. The lower part of the walls around is
made of concrete. The materials for the upper part are, however,
quite different: plastic in the
north and concrete for East and South walls with 11\,cm thick metal
pillars distributed every approximately 2.5\,m all around. North and East walls faced by the TPC detector
during magnet movement are shown in  Figure \ref{Site}.

 The inhomogeneity of the
building materials led us to undertake a careful study of the radioactive background and a detailed analysis of the measured
TPC background data.

\subsection{Gamma background\label{subsec:gamma}}

 An hyper pure germanium gamma spectrometer system, ISOCS
(In-Situ Object Counting System), based on a HpGe coaxial detector
from Canberra has been used for gamma ray measurements in a range from
50\,keV to 3\,MeV \cite{Tech}.
 The gamma spectrometry measurements confirmed the radioactive chains and
potassium as the main sources for background and showed a background
disparity between the different types of walls (see Table
\ref{gmeas}).
\begin{table}[h]
  \centering
  \caption{Mean gamma production in the CAST site [Bq/kg]. For the radioactive chains, equilibrium activities are quoted. In the case of radon emanation from
  the $^{238}$U chain, equilibrium is broken and activities for  the nuclides before and after $^{222}$Rn are given separately.  }\label{gmeas}
  \vskip 0.5 cm
   \begin{indented}
  \item[]\begin{tabular}{@{}c c c c c c}
  \br
 Wall& \multicolumn{2}{c}{$^{238}$U Chain} &$^{235}$U & $^{232}$Th & $^{40}$K  \\
description &  $^{238}$U$\rightarrow ^{226}$Ra& $^{218}$Po$\rightarrow
^{210}$Po & chain & chain&  \\ \mr
  East (lower) &  \multicolumn{2}{c}{$25 \pm 2$} &
$1.1\pm 0.7$&$10\pm 2$& $113\pm10$\\
and North Pillars & & & & & \\
  East (upper) & 923$\pm$274 & 32$\pm$5 & 40$\pm$12 &
$34\pm6$ & $388\pm45$ \\
and South (upper) & & & & & \\
 \br
  \end{tabular}
    \end{indented}
\end{table}

Assuming a concrete density around $2.4\,\rm{g\,cm}^{-3}$ and a wall thickness of
30\,cm, we estimate the gamma production in the lower part of
every wall as $\sim$ 2.25 photons $\rm{cm}^{-2}\,\rm{s}^{-1}$, around
8 photons $\rm{cm}^{-2}\,\rm{s}^{-1}$ in the upper part of the east
and south walls and $\sim$ 2 photons $\rm{cm}^{-2}\,\rm{s}^{-1}$ in
the north wall pillars. We should also add around 6 photons
$\rm{cm}^{-2}\,\rm{s}^{-1}$ coming from the soil.

These data pointed also to a radon emanation for the east and south
walls. Later radon measurements \cite{Tech} gave mean values of $20\pm10\,\rm{Bq/m}^3$ in summer and
$15\pm5\,\rm{Bq/m}^3$ in winter, not incompatible with a certain dependence on temperature.

Within this category, we could also include gammas arriving from
radioactive contamination in the magnet platform, electronics
material and also the contribution of the detector itself. Most of
these materials are steel, plastics and metals (i.e. Fe, C, H, Cu,
...) whose impurities are also uranium, thorium and potassium.

  Neutrons and protons coming from cosmic rays
can also induce radioisotopes in the detector gas and materials
(mainly $^{14}$C and $^3$H) whose contribution can be neglected
owing to their low production rate (0.003 nuclei of $^{14}$C and
0.001 nuclei of $^3$H per litre and day in argon, computed
considering saturation with a modified version of the COSMO
\cite{COSMO} code based on the semiempirical formulas of Silberberg
and Tsao \cite{ST}).  Another gamma contribution corresponds to
the cosmic ray photon flux, being only a small fraction ($\le 1\%$) of
the total \cite{Heusser}.

Summarizing, the experimental site contributes to an important, and non-uniform, gamma
 background owing to radioactive contamination.
  Most of the gamma
 radiation described above would traverse the active volume of the detector
 without interacting at all due to the special features of the
 detector
 (only sensitive to energies of a few keV).
  However, energetic gamma background loses part of its
 energy after interactions with materials surrounding the detector creating
 secondary photons which do contribute significantly to
 the  background signal. Therefore, any shielding designed for the
 TPC detector should be a compromise between the external flux reduction and
 the increase of interactions in these materials as well as their
 own radioactive contamination. The best control of all these
 variables is achieved after Monte Carlo simulations plus
 experimental tests.

\subsection{Neutron background\label{subsec:neutrons}}

Even if the neutron component of the background is below the level
of the typical gamma background by three or four orders of
magnitude, neutron signals in the detector could mimic those from
X-rays.

Quantitative measurements of neutron background have been performed
in the experimental site with a BF$_3$ detector.
 The homogeneous measured flux of neutrons in the CAST site is around
 $3\times 10^{-2}\,\rm{cm}^{-2}\,\rm{s}^{-1}$. This value, and its homogeneity, points to a cosmic source.
 Cosmic ray generated neutrons have  energies below a few
 GeV and the spectrum shows a dependence as 1/E$^{0.88}$
 up to 50\,MeV and as 1/E above this energy \cite{Hess}. This is the most
 important neutron contribution, not only for its intensity but
 also for its high energy.  Other sources of neutron
background are
 neutrons induced by muons in the surrounding materials,
$(\alpha,n)$ reactions on light elements and the spontaneous
fission.

Muons interacting in shielding materials produce neutrons. FLUKA
\cite{fluka} simulations with a measured total muon flux of $\sim$
$50\times 10^{-3}\,\rm{cm}^{-2}\,\rm{s}^{-1}$ gave us a yield for
neutrons entering the detector of $1.2\times
10^{-3}\,\rm{cm}^{-2}\,\rm{s}^{-1}$ and a neutron spectrum peaking
below 1\,MeV. Most of these events are rejected by anti-coincidence with the
veto installed as part of the shielding.

In nature mainly three nuclides ($^{238}$U, $^{235}$U and
$^{232}$Th) undergo spontaneous fission. The rate of spontaneous
fission of $^{238}$U, the nuclide of shorter half life, is
0.218/year/g of concrete for 1\,ppm of $^{238}$U and the average
number of neutrons emitted per fission event is $2.4\pm 0.2$ with a
typical evaporation spectrum peaking at around 1\,MeV. Most of these fission
neutrons will come from the concrete walls. Assuming a penetration for neutrons of 10\,cm, we
 estimate the volume and surface production of neutrons in the
concrete walls from around 0.60$\times
10^{-6}\,\rm{cm}^{-2}\,\rm{s}^{-1}$ for the lower East wall and $1
\times 10^{-6}~\,\rm{cm}^{-2}\,\rm{s}^{-1}$ for the metal pillars to
30 $\times 10^{-6}\,\rm{cm}^{-2}\,\rm{s}^{-1}$ in the upper East and
South walls.

The $\alpha$ particles emitted in the radioactive chains can
interact with other elements and produce neutrons through
$(\alpha,n)$ reactions. For the upper South and East
walls the final estimated
neutron yield , following \cite{Wulandari}, is $ \sim 10^{-5}\,\rm{cm}^{-2}\,\rm{s}^{-1}$. The
radioactive activity of soil (or of the lower East wall) will
produce one order of magnitude less of neutrons $\sim
10^{-6}\,\rm{cm}^{-2}\,\rm{s}^{-1}$. The energy spectra
for these neutrons consist of peaks related to $\alpha$ particles
energies, with 8.79\,MeV being the highest energy for naturally
emitted $\alpha$ particles (decay of $^{212}$Po).

In summary, we have collected all the neutron productions in Table
~\ref{tneutrons}. The energetic cosmic component appears as the most
relevant contribution.
\begin{table}[h]
  \centering
  \caption{Comparison of the estimated order of magnitude for neutrons coming from different sources.
  Values are given in neutrons per cm$^{2}$ and second.}\label{tneutrons}
  \vskip 0.5 cm
   \begin{indented}
  \item[]\begin{tabular}{@{} c c c c}
   \br
  Cosmic  &Muon induced & Fission  &$(\alpha,n)$  \\
    \mr
    $\sim 10^{-2}$ &
    $\sim 10^{-3}$& $\sim 10^{-5}$ & $\sim 10^{-5}$
   \\
\br
  \end{tabular}
    \end{indented}
\end{table}

\section{Monte Carlo simulations \label{sec:simulations}}

A complete set of Monte Carlo simulations has been carried out in
order to estimate the external background contribution to the TPC detector data during the 2003 and 2004 taking data
periods. We have not only reproduced the TPC detector and its shielding in the simulations, but also the main software cuts to be able to discriminate the expected
X-ray signal. They show that the use of cuts reduce the registered events
by two orders of magnitude in good agreement with the experimental data. This fact results in low statistics, having to establish a compromise between computing time and
non-negligible statistical errors. Despite this, Monte Carlo
simulations have been very useful to understand and quantify the
number of counts coming from the external background in the 2003 and
2004 sets of data of the TPC detector.

 The
background level (in the 3--7\,keV region of interest) for three different shielding configurations  have been compared: a 5\,mm-thick copper box, the copper
box plus 2.5\,cm of lead, and a complete shielding
consisting of 5\,mm-thick copper box
 plus 2.5\,cm of lead plus 22.5\,cm of polyethylene.

First of all we will focus on gamma simulations since this is the
main contribution to the TPC background, then we will make a few
comments about neutron simulations, though they are quite difficult
to verify experimentally.

\subsection{Simulations for external gamma\label{subsec:gamma-sim}}

These simulations have been performed using the GEANT4 code
\cite{geant4} since it allows to take into account all the aspects of the
simulation process including the detector response. The primary events, corresponding to the
radioactive chains and potassium, have been generated uniformly and
isotropically on a sphere surrounding the most external surface of
the shielding.

The shielding made of 5\,mm of copper plus 2.5\,cm of lead plus
22.5\,cm of polyethylene reduces the external gamma background by
more than one order of magnitude, (92$\pm$3)$\%$ in the 3--7\,keV
range. Since the thick layer of polyethylene helps in the gamma attenuation,
the same shielding without polyethylene is about a 15$\%$
less effective, causing an estimated reduction of (77$\pm$4)$\%$. Compton interactions in the polyethylene
result in lower energy photons which are easily absorbed in the lead.
Though a thicker layer of lead could stop a larger fraction of the external
 gammas, it
can be a source of secondary neutrons and it would
add too much weight to one of the ends of the magnet.

Due to the spatial inhomogeneity of the gamma background coming from walls, simulations allow us to make just a
rough estimate of the TPC recorded counts. These external photons
cause between 30 and 55 counts per hour in the volume of the
detector facing the two windows
 for the 3--7\,keV energy region in the case of the copper shielding (2003 data),
  and between 2 and 5 counts per hour in the case of the complete shielding configuration of 2004.
  This result, as we will see in next Section, is
  compatible with measured data.

   The GEANT4 package has also been used to simulate the effect of the radon trapped inside the copper box.
   Radon decays have been produced isotropically in the volume between the TPC and the copper
   box.
  These simulations
  quantify the point-like signals in the 3--7\,keV range in 0.04 counts per hour for an overestimated radon concentration of
  25\,Bq/m$^3$. Therefore, the radon contribution to the background can be
 neglected in the 2003 and 2004 shielding conditions due to the plastic
 bag around the entire set-up and the nitrogen flush.

\subsection{Neutron simulations\label{subsec:neutron-sim}}

Though neutrons interacting in materials can produce $\gamma$
particles, more neutrons, $\alpha$ particles and fission fragments
depending on materials and energies, the most efficient reaction is
the elastic scattering, energy transferred to nuclear recoils. This
energy  is determined by the energy of the neutron incident ($E_n$)
and the scattering angle $\theta$
\begin{equation}\label{erecoil}
    E_R=\frac{4A}{(1+A)^2}(\cos^2\theta) E_n.
\end{equation}
In the case of argon, $A=40$, assuming  a quenching factor of 0.28,
the maximum visible energy and the neutron energy are related as
follows
\begin{equation}\label{evisible}
    E_{Rmax}=0.0266 E_n.
\end{equation}
Therefore, the neutrons able to deposit a visible energy in the
analysed  range of 3--7 keV are mostly those with an energy between
0.11 and 0.27 MeV.

A GEANT4 simulation, using G4NDL3.5 high
precision neutron data library and a real cosmic spectrum as input,
has been carried out. This simulation has allowed us to roughly
estimate about 2 counts per hour caused by cosmic neutrons inside
the two windows of the TPC in the (3--7)\,keV visible energy region
in the case of the copper shielding. As we will show in
Section~\ref{sec:tpc-measurements}, this rate is between 25-45 times
smaller than the counting rates in 2003.

\begin{figure}[ht]
\begin{center}
\hspace{0 cm} \psfig{figure=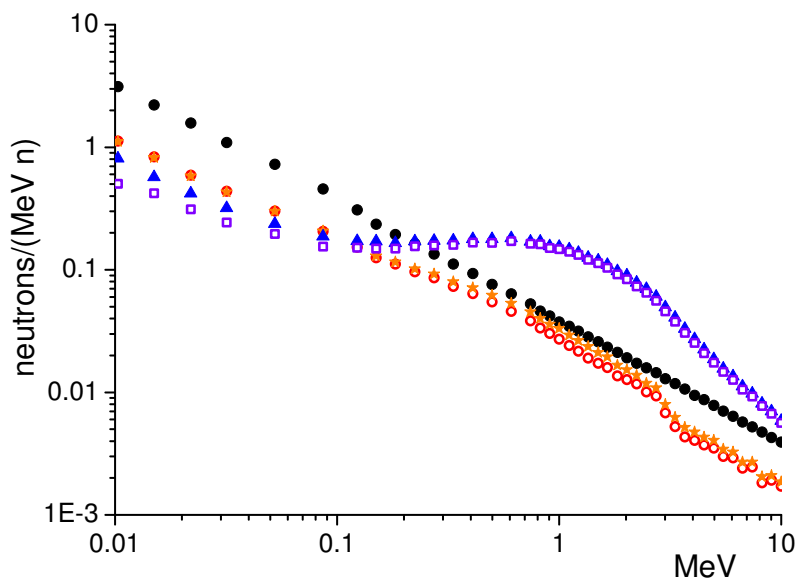,width=120mm} \caption{Cosmic
neutrons in the 10 keV-20 MeV range after traversing every layer of
shielding: incoming cosmic neutrons (solid circles), neutrons after
traversing 22.5\,cm of polyethylene (open circles) plus 1\,mm of
cadmium (solid stars) plus 2.5\,cm of lead (solid triangles) plus
5\,mm of copper (open squares). Spectra are normalised to one cosmic
neutron. \label{cfneutrons}}
\end{center}
\end{figure}

 To understand the effects of
the different layers of shielding we have used the FLUKA code of
proven reliability for the transport of neutrons.  This second
simulation shows the effects of every layer of shielding material on
cosmic neutrons: while polyethylene decreases the number of
background neutrons per cosmic neutron, the 2.5\,cm of lead
increases this number (see Figure~\ref{cfneutrons}) due to $(n,2n)$
processes (observe the evaporation spectrum). Cadmium absorbs
thermal and epithermal neutrons with energies below 1keV.

 The shielding without polyethylene  has been also compared to the complete
configuration. The number of cosmic neutrons able to produce nuclear
recoils and a deposit of visible energy in the TPC in the 3--7\,keV
range decreases by only 20$\%$ after the complete shielding due to
the production of neutrons in lead. The number of neutrons could
even  increase by 10$\%$  if the 22.5\,cm of polyethylene are taken
off. Then, the shielding hardly reduces the amount of neutrons and
it would increase the number  without polyethylene. This background contribution has been
estimated in around 1.5 counts per hour for the 2004 shielding conditions.

The neutron production in the shielding due to muons has also been
investigated. Most of these neutrons are produced in lead. As
mentioned in Subsection~\ref{subsec:neutrons}, the number of
neutrons produced by each sea level muon is 0.024, giving a neutron
flux reaching the detector of 1.2$\times 10^{-3}$ per
$\rm{cm}^{-2}\,\rm{s}^{-1}$, six orders of magnitude higher that the
calculated in \cite{canfranc} for a thicker shielding but a much
less intense underground muon flux. We can roughly estimate about
0.02 the counts per hour inside the windows corresponding to
neutrons induced by muons, three orders of magnitude smaller than
the measured background rates.

Regarding other population of neutrons, we have not considered those
coming from radioactivity (fission and ($\alpha$, n) processes)
since 22.5\,cm of moderator would reduce their tiny contribution (of
the order of $10^{-5}\,\rm{cm}^{-2}\,\rm{s}^{-1}$) by two orders of
magnitude \cite{canfranc}.

\section{TPC  background data\label{sec:tpc-measurements}}

First tests of the full shielding (with a 5-cm-thick layer of lead in this case) at the laboratory showed a reduction
factor of $\sim 8$ below the background level of the chamber, without any shielding or nitrogen flush, for energies between 1 and 10\,keV. Once the
detector was mounted in the magnet on the moving platform, this
factor became $\sim 4.3$ ($\sim 6.4$ in the 6-10\,keV range).

\begin{figure}[ht]
\begin{center}
\hspace{0 cm} \psfig{figure=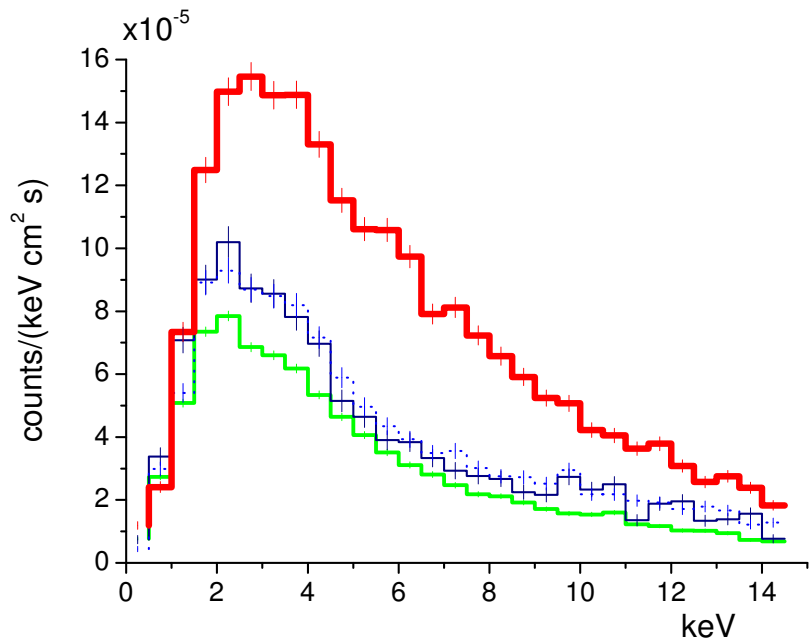,width=120mm}
 \caption{Background data obtained in the shielding test at the experimental site for different
 shielding configurations: full shielding set-up (bottom line), the
 copper box plus 2.5\,cm of lead (solid line in the middle), the
 copper box plus a double layer of lead (dotted line in the middle)
 and just the copper box (upper line).
\label{test}}
\end{center}
\end{figure}

After these first tests, another set  was undertaken at the CAST
experimental site to observe the effects of every
component of the shielding. The magnet was placed in horizontal position
facing the NE corner, far away from the forced air windows and in an
intermediate position between the North and the East wall in order
to get a more \emph{average} flux. Here we measured the TPC
background in different shielding conditions. For this position, it
has been checked that the full shielding (copper plus lead plus
polyethylene) reduces background levels by a factor of $\sim$3 in the 3--7 keV
energy interval, in comparison to the copper shielding, thanks
mainly to the 2.5\,cm of lead, while a double layer of lead (5\,cm)
does not improve these results (see Figure~\ref{test}). When
comparing the experimental reduction factor from those obtained in
simulations, it must be kept in mind that in the latter we were just
dealing with external gamma radiation, forgetting about internal
contamination or radon intrusion which contribute to measurements in
all shielding conditions. Other reason for the discrepancies lies in
the fact that in simulations the shielding is all around the
detector while the TPC is actually attached to the magnet pipes, and
thus partially not shielded from it. As a consequence of these two reasons,
the background level decrease is lower than expected after
simulations.

 Other experimental test was performed to quantify the contribution of the radon trapped in the
  copper box. Measurements were carried out with and without
  nitrogen flush at the same spatial position and one right after the
  other to avoid time variations. The subtracted spectrum
  (without and with nitrogen flush) can be
   thought as due to the radon inside the copper box since the plastic bag prevents the
  outer radon from entering the shielding. Despite the poor statistics, the estimated radon
  rate in the 3--7\,keV range for point-like signals (0.13$\pm$
  2.33
  counts per hour) in the volume facing the two windows of the TPC,
  shows a negligible contribution to the background and is compatible with the 0.04 counts estimated in simulations.

\begin{figure}[ht]
\begin{center}
\hspace{0 cm} \psfig{figure=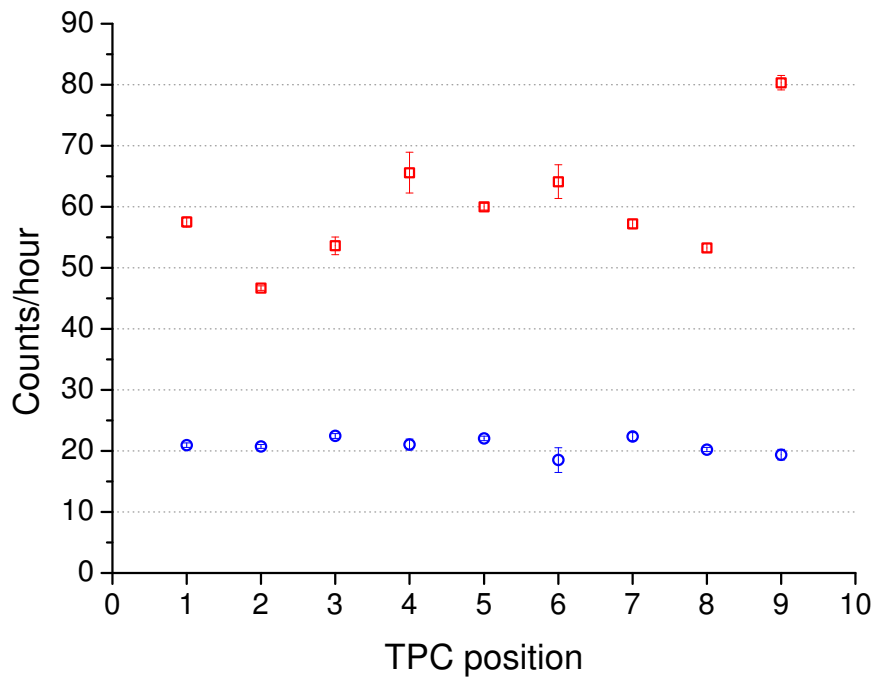,width=130mm}
 \caption{\label{fondos}July-August background data for 2003 (squares) and 2004 (circles).
Measurements correspond to 3 vertical positions and 3 horizontal
positions of the TPC detector. The 9 magnet positions run first vertically
and then horizontally.}
\end{center}
\end{figure}

Finally, we can also compare the experimental background data, since during the two
data taking periods (years 2003 and 2004) the TPC detector
has recorded not only tracking data, when the opposite part of the
magnet is pointing to the sun, but also background data at any other
time of the day. In order to get the best control of the background and determine its inhomogeneity, these measurements have been
performed at precise horizontal (movement along a circle) and
vertical positions. The Figure \ref{fondos} shows measured background levels for
  nine positions: the three first measurement points are facing
the North wall; the three next  points  face the NE corner and the
last one faces the East wall near one of the metal pillars. During the year 2003, with the TPC covered by a 5\,mm thick copper box
 and a nitrogen flush inside, the background measurements showed a high
 degree of inhomogeneity as it can be observed. Higher background rates  are registered in the
 proximity of more intense sources of radioactivity such as the upper part of the
 East wall or the
 soil
 while the closeness to a metal pillar or to the plastic wall
 (upper part of North wall), decreases rates.
  Also  metal components as scaffolding, ladders,...
can affect measurements at some points. Thanks to the shielding
described in Section~\ref{sec:tpc}, the 2004 background data show
rate reductions by more than a factor of 2.5, reaching even a factor of 4
at some positions (see Figure~\ref{fondos}). Moreover, the
background is now fairly homogeneous.

When comparing these experimental data with the results of the simulations presented in the previous section, one notes than the background reduction obtained with the shielding is larger in the external gamma simulations (a reduction factor of 10 -- 15) than in the experimental data (2.5 -- 4). This seems to imply that, while most of the background in the 2003 setup was indeed linked to the external gamma background (as proved by its inhomogeneity and the effectiveness of the shielding), most of there remaining background in the 2004 setup must be however of a distinct origin that those included in the simulations. The most probable candidates are internal contaminations of the components inside the shielding, or background coming from the side where the detector is attached to the magnet (and necessarily unshielded), from contaminations present for instance in the magnet components. This hypothesis fits well with the fact that this contribution in constant for every magnet position. As shown in Table \ref{btable}, where all calculated and measured background levels are collected, a speculated constant internal contamination (first column) of about 15 counts per hour would fit the overall scenario. Whatever the precise origin of the remaining background, the effect of the shielding on the 2005 TPC operation, both in terms of reduction of background and its variability, yielded a substantial increase of sensitivity of the detector in the context of the CAST experiment \cite{CAST2}, when compared to the previous 2003 period.

\begin{table}[h]
  \centering
  \caption{Estimated external background contribution to TPC data corresponding to the years 2003 and 2004.
  Values are given in counts per hour and correspond to estimated point-like, 3--7 keV energy, deposits in
  the TPC volume facing the two windows of the chamber, compared to the mean measured values}\label{btable}
   \vskip 0.5 cm
  \begin{indented}
  \item[]\begin{tabular}{@{} c c c c c c c}
   \br & \multicolumn{5}{c} {Monte Carlo estimates} & Measured\\
 Year & Internal & External $\gamma$  & Cosmic & $\mu-$induced & Radon  & values\\
  & cont. & sources & neutrons & neutrons &   &  \\\mr
  2003 & 15 & 30-55 & 2 &  &0.04 &47-80\\
  2004 & 15 & 2-5 & 1.5 & 0.02 &0.04 & 19-21
    \\
\br
  \end{tabular}
    \end{indented}
\end{table}

\section{Summary and conclusions \label{sec:conclusion}}

After a characterisation of the radioactive contamination, we
present herein the effects of a shielding on the CAST TPC detector
background data. Requirements of a reduction of the background
levels and of a decrease of the background inhomogeneity have been
fulfilled.

 Gamma measurements have reported a clear inhomogeneus
radioactive contamination due to the uranium and thorium radioactive
chains and to potassium and radon emanation from the East and South
wall. Also neutrons have been studied, being the cosmic neutrons
the most relevant contribution. Other
background contaminants can be neglected.

Then, after the identification of the background sources, we have undertaken Monte
Carlo simulations which have allowed us to understand the TPC detector
response to gamma and neutron sources as well as the effects of
different components of shielding on the background levels. As a result
of these simulations, we have learned that 2.5\,cm of lead reduces
the external gamma background in the 3--7\,keV range by $(77\pm 4
)\%$ but produces neutrons due to high energy cosmic neutron and
muon interactions. The addition of a 22.5\,cm layer of polyethylene
results in a $(92\pm 3 )\%$ reduction of the gamma background, decreases by $20\%$ the
cosmic neutrons and eliminates any low energy neutrons. These results are compatible with experimental tests
performed in the CAST site. For all these reasons the installed
shielding in 2004 consists of 5\,mm of copper plus 2.5\,cm of lead
plus 22.5\,cm of polyethylene.

Finally, the 2004 data confirm a reduction of background levels by a
factor between 2.5 and 4 from the 2003 data (the highest for
positions close to the most intense gamma sources) as well as a
quite acceptable degree of homogeneity. For this period, most of the
contamination is due to sources near the detector (around 15 counts per hour
in the volume facing the two windows for the 3--7\,keV energy
interval), while external gamma radiation and cosmic neutrons only
add 3-6 counts per hour to the TPC detector  background for the same interval of energies.

\ack

This research was partially funded by the Spanish Ministry of
 Education and Science  (MEC) under contract FPA2004-00973 and
  promoted within the  ILIAS
(Integrated Large Infrastructures for Astroparticle Science) project
funded by the EU under contract EU-RII3-CT-2003-506222. We thank the
CAST Collaboration for their support. Our gratitude also to the
group of the Laboratorio Subterr\'{a}neo de Canfranc (LSC) for material
radiopurity measurements.


\section*{References}

\begin{thebibliography}{21}
\bibitem{CAST1} Zioutas K  \etal 1999 \NIM A {\bf 425}  480
\bibitem{CASTTPC}  Autiero D \etal 2007 {\it New Journal of
Physics (in press)} ({\it Preprint: physics/0702189})
\bibitem{CCDT}  Kuster M \etal 2007 {\it New Journal of
Physics (in press)} ({\it Preprint: physics/0702188})

\bibitem{teles} Lutz G \etal 2004 {\it Nucl. Instr. and Meth} A {\bf 518}
201
\bibitem{MMT} Abbon P \etal 2007 {\it  New Journal of
Physics (in press) } ({\it Preprint: physics/0702190})

\bibitem{Castpaper}Zioutas K \etal [CAST Collaboration] 2005 {\it Phys.Rev.Lett.} {\bf 94} (2005)
121301.


\bibitem{CAST2} Andriamonje S \etal [CAST Collaboration] 2007 {\it JCAP journal}, {\bf 04} 010, ({\it Preprint: hep-ex/0702006})




\bibitem{TAUP}  Ruz J \etal 2006 {\it  Proc. of the 9th International Conference on Topics in
Astroparticle and Underground Physics 2005 (TAUP05), Journal of
Physics: Conference Series} {\bf 39} 191

 \bibitem{Tech} CAST Technical paper, in preparation.
 \bibitem{COSMO}
 Martoff C J \etal 1992 {\it Comp. Phy. Commun.} {\bf 72} 96

 \bibitem{ST}
Silberberg R and Tsao C H 1973 {\it Astrophys. J. Suppl. Ser.} {\bf
25} 315, ibid p. 335.

 \bibitem{Heusser}
 Heusser G 1995 {\it Annu. Rev. Nucl. Part. Sci.} {\bf 45} 543

\bibitem{Hess}
Hess W N \etal 1959 {\it Phys. Rev.} {\bf 116} 445



\bibitem{fluka}
Fass\`{o} A \etal 2000 {\it Proceedings of the MonteCarlo 2000
Conference (Lisbon)} Eds. Kling A, Barao F, Nakagawa M, Tavora L and
Vaz F ( Berlin: Springer-Verlag Berlin) p.~159 (2001); Fass\`{o} A
\etal ibid, p.~955




\bibitem{Wulandari}
Wulandari H \etal 2004 {\it Astropart. Phys.} {\bf 22} 313




\bibitem{geant4}
Agostinelli S \etal [GEANT4 Collaboration] 2003 \NIM A {\bf 506} 250



\bibitem{canfranc} Carmona J M \etal 2004 {\it Astrop. Phys.} {\bf 21} 523
\end{thebibliography}
\end{document}